\documentclass[runningheads,a4paper]{iaga}

\usepackage{amssymb}
\usepackage{graphicx}

\usepackage{url}
\newcommand{\keywords}[1]{\par\addvspace\baselineskip
\noindent\keywordname\enspace\ignorespaces#1}

\begin{document}

\mainmatter  

\title{Formation of the current sheet in a coronal streamer}

\titlerunning{Formation of current sheet in coronal streamer}

\author{Lucia Abbo\inst{1}, Ester Antonucci\inst{1}
Roberto Lionello\inst{2}, Zoran Miki\'c\inst{2}, Pete Riley\inst{2}}
\authorrunning{Abbo L. et al.}

\institute{INAF-Astronomical Observatory of Turin, 10025, Pino Torinese, Italy,\\
\email{abbo@oato.inaf.it}
\and
Predictive Science Inc., San Diego, CA 92121, USA}

\toctitle{Formation of current sheet in coronal streamer}
\tocauthor{Abbo L. et al.}
\maketitle

\begin{abstract}
The present work is on the study of a coronal streamer observed in March 2008
 at high spectral and 
spatial resolution by the Ultraviolet Coronagraph Spectrometer (UVCS)
 onboard SOHO.
On the basis of a spectroscopic
analysis of the O VI doublet, 
the solar wind plasma parameters are inferred
in the extended corona. The analysis accounts for the coronal magnetic topology, 
extrapolated
through a 3D magneto-hydrodynamic model. 
The results of the analysis show indications
on the formation of the current sheet, one of the source regions of the slow coronal wind.
\end{abstract}

\section{Introduction}
The origin of the slow solar wind is one of the open problems
 in solar physics. During solar minimum, the slow solar wind tends to be confined in the equatorial and mid-latitude regions as confirmed by the Ulysses observations in the heliosphere. The main issue is whether the slow wind is coming from open field line regions surrounding the streamer
or there is a substantial contribution from the streamer itself: some authors proposed that
 the low-speed wind could flow between structures present in the inner part
of the large equatorial streamers (Noci et al. 1997; Noci \& Gavryuseva 2007); for others, the slow wind arises from the streamer bright regions observed in UV identified with streamer stalks (e.g. Habbal et al. 1997, Raymond et al. 1997, Strachan et al. 2002, Uzzo et al. 2003); other papers indicate the open flux tubes in the streamer adjacent regions as the main slow wind sources (e.g. Abbo et al. 2003, Antonucci et al. 2005, Abbo et al. 2010). 
The analysis presented here concerns data from the recent solar minimum (2006-2008) which has been very peculiar and different in comparison with the previous one (Gibson et al. 2009). In order to investigate signatures of the slow wind sources, we derive the HI and OVI kinetic temperature from the data analysis of the spectral lines observed by UVCS/SOHO and the coronal electron density as a function of the outflow velocity through a dignostic techniques described in the next session.

\section{Diagnostic techniques for the coronal plasma}
Outflow velocity and electron density of the coronal wind plasma can be deduced from the
emission of
intense ultraviolet spectral lines, such as the O~VI 1032, 1038 lines. These lines are formed
in the extended corona via collisional
 and radiative excitation
processes. The two components have a
different dependence on the electron density: the collisional process depends
on $n_e^2$, while the radiative process
  depends linearly on electron density $n_e$.
The collisional and radiative components of the O~VI 1032, 1038 lines in an expanding plasma can be separated by
using the method introduced by Antonucci et al. (2004).
\begin{figure}
\centering
\includegraphics[height=5cm]{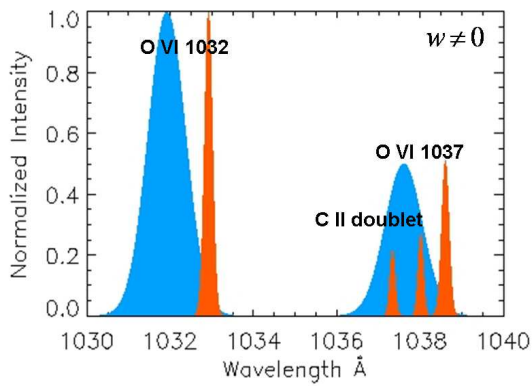}
\caption{Scheme of the Doppler dimming effect for the doublet OVI lines. The absorption profiles of the 1032 and 1037 OVI lines are shown in blue; the exciting profiles of the two doublet OVI and CII lines are plotted in red, shifted for Doppler effect (w$\neq$0). The absorption profiles are larger because the temperature is higher in corona than in the lower solar atmosphere.}
\label{fig1}
\end{figure}
The electron
density, averaged along the line-of-sight (l.o.s.), $<n_e>$, is proportional to the ratio of the collisional component, $I_c$,
 to the radiative component, $I_r$, and is a function of the outflow
velocity of the wind, {\bf w}:
$<n_e>\,\sim\,\frac{I_c}{I_r}\,<\Phi(\delta\lambda)>$,
where
$<\Phi(\delta\lambda)>$ is the Doppler dimming function which  depends
on the normalized
 coronal absorption profile and on the intensity of the exciting spectrum along the direction of the incident radiation,
$\bf n$. The quantity $\delta\lambda=\frac{\lambda_0}{c}\,{\bf w}\cdot{\bf n}$ is
 the shift of the disk spectrum
introduced by the expansion velocity, {\bf w}, of the coronal absorbing
ions/atoms along the direction $\bf n$ and $\lambda_0$ is the reference wavelength of the
transition. As the wavelength shift increases, the resonantly scattered emission decreases,
 giving origin to the Doppler dimming effect (Beckers 1974; Noci et al. 1987). By analysing the
 O~VI doublet lines at 1031.93 and
1037.62 \AA, it is
possible to measure oxygen ion outflow velocities up to approximately  450 km s$^{-1}$ for
 the effect of pumping
of the CII lines at 1037.02 and 1036.34 \AA~on the O~VI $\lambda$~1037.61 line (e.g.
 Dodero et al., 1998).\\
When the plasma is dynamic, in order to measure the coronal electron
 density and the outflow velocity
at the same time, we impose the constraint of the mass flux conservation along the
flow tube connecting the corona to the heliosphere by taking into account the expansion factors of the flux tubes as derived by the MHD model and by considering the mass flux measured in the heliosphere
by Ulysses (McComas et al. 2008).

\section{Extrapolations of the coronal magnetic field}
In order to get a detailed description of the magnetic topology in the outer corona and derive the magnetic field expansion factors of the flow tubes,
 the coronal magnetic fields have
been extrapolated from photospheric longitudinal fields on the basis of
the three-dimensional MHD model of Miki\'c et al. (1999).
The code integrates the time-dependent MHD equations in spherical
 coordinates ($r,\theta,\phi$) until the plasma and magnetic fields settle
 into equilibrium.
The
photospheric magnetic field data (obtained from synoptic observations at
 Kitt Peak National Solar Observatory on the days of observation considered in the analysis)
 are used to specify the boundary condition on the
radial component of the magnetic field, $B_r$. Other boundary conditions on the velocity, temperature and density of plasma are determined
 with constraint values at the solar surface.

\section{Observations and data analysis}
We have analyzed UVCS observations of a streamer performed in the period 14-20 March 2008,
characterized by a good spatial  coverage (heliodistances of the slit center from 1.7 to 3.5 R$_\odot$).\\
\begin{figure}
\centering
\includegraphics[height=5cm]{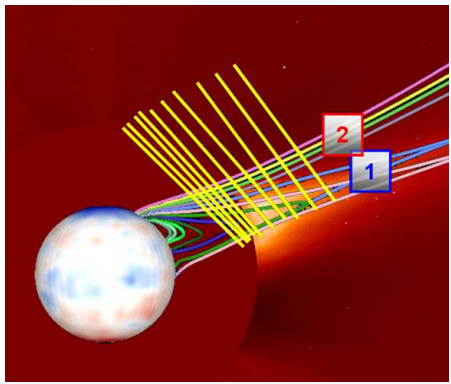}
\caption{Composite image: visible light in corona from LASCO C2, magnetic field lines as derived by yhe MHD model and UVCS field of view (yellow lines). }
\label{fig2}
\end{figure}

The pointing of UVCS was carefully studied, in order to center the cusp of the streamer in the part of the slit which still has the maximum spatial resolution after the OVI detector's electronic problems since January 2006. 
For this observation, the spatial resolution is 28 arcsec and the slit width corresponds to 100 $\mu$m up to 2 R$_\odot$, to 150 $\mu$m from 2.1 to 2.5 R$_\odot$ and to 200 
$\mu$m from 2.7 to 3.5 R$_\odot$. 
The magnetic field line map is shown in Figure 2 overlapped with the coronal image in visible light from LASCO C2 and the field of view of UVCS
 indicated by yellow lines. The streamer boundary is derived on the basis of the MHD model and divided two regions considered in the analysis: region 1 characterized by closed magnetic toplogy and region 2 with open magnetic field lines (labeled in Fig.2) . 
  In order to derive the plasma conditions within the streamer and in the external region,
the intensities of the O~VI doublet lines are integrated in the two regions by applying the radiometric calibration of UVCS data.
The electron density and the outflow velocity are derived from the ratio of the
collisional
to radiative component of the oxygen O~VI 1032 line,
with the constraint of mass flux conservation, according to the method discussed in the previous section.
The expansion factors of the flux tubes connecting the corona and heliosphere, are
derived from the extrapolations of the coronal magnetic fields of the MHD model.

\section{Results and conclusions}
We would like to study the variation of the coronal plasma physical parameters
 in open and closed magnetic field lines regions in order to recognize possible signatures of the slow wind sources and of the current sheet formation.
The kinetic temperature of ions, expressed in terms of the spectral line width observed by UVCS,  $T_k \propto \sigma_\lambda^2$,
 is a measure of the velocity distribution of the ions
along the line of sight.
\begin{figure}
\includegraphics[height=4.4cm]{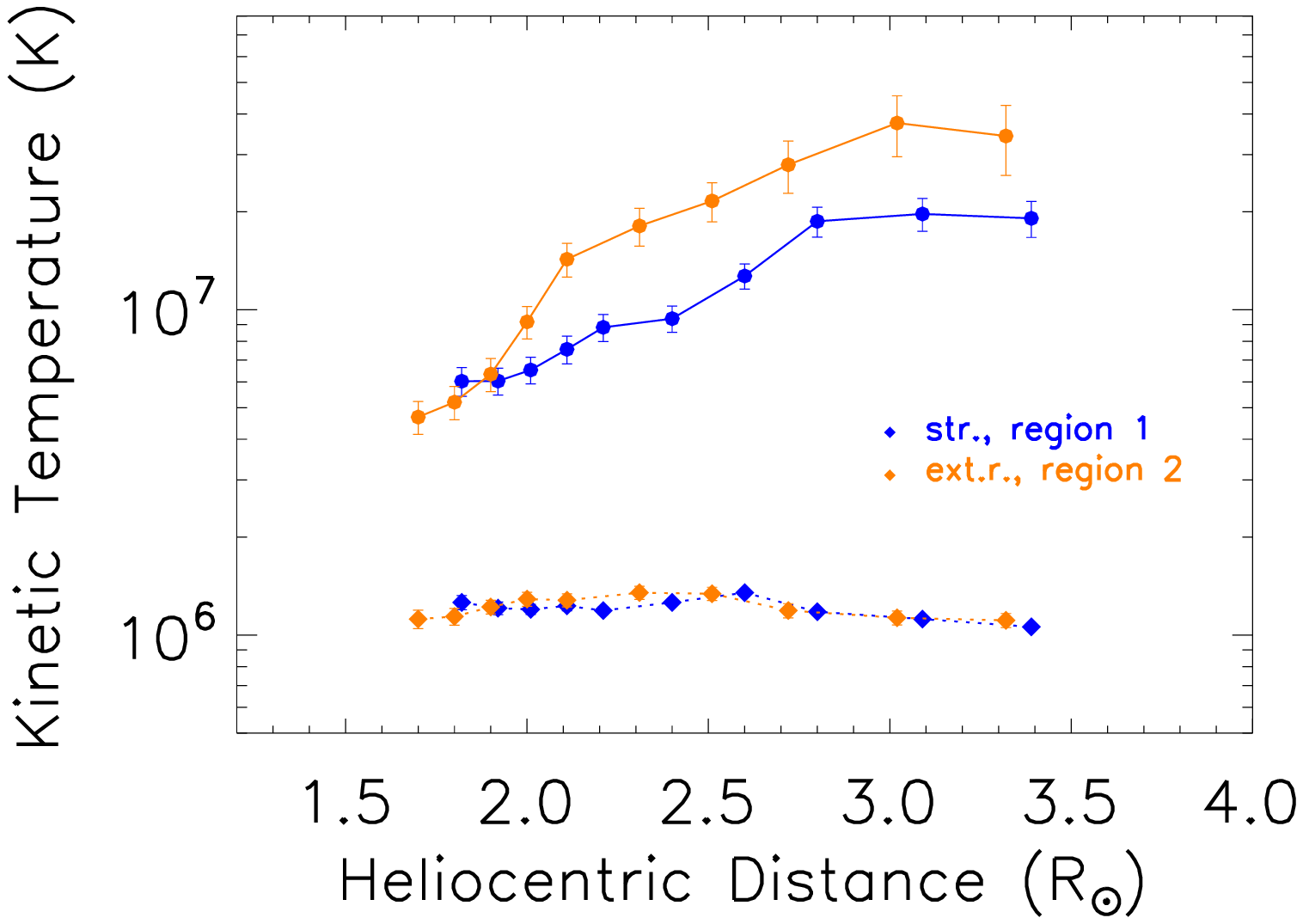}
\includegraphics[height=4.4cm]{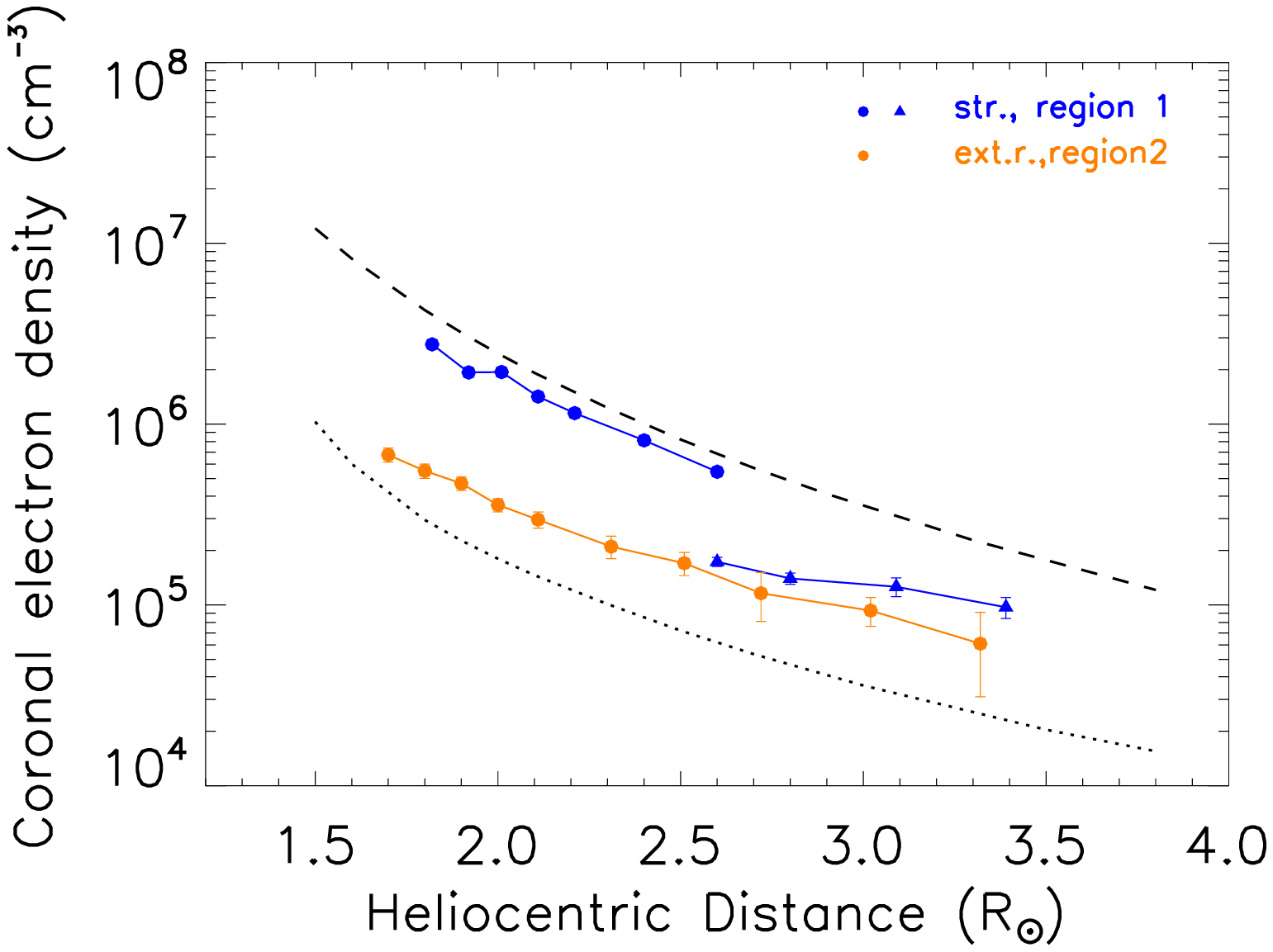}
\caption{Left panel: Kinetic temperature of HI Ly$\alpha$ (diamonds) and OVI 1032 (dots) as a function of heliodistance for the two regions considered in the analysis. Right panel: Electron density as a function of heliodistance relative to the internal part of the streamer (region 1) and the adjacent part (region 2). The dashed line shows the values derived by Gibson et~al. (1999) from visible light observations for streamers and the dotted line shows results obtained by Guhathakurta et~al. (1999) for coronal holes.}
\label{fig3}
\end{figure}
 The left panel of Figure 3 shows the kinetic temperature of HI atoms (diamonds) and OVI ions (dots) as a function of heliodistance for the two regions defined in Fig.1 (blue points for region 1 and orange for region 2). It is worth to note that, on the one hand, the $T_k$ of HI Ly$\alpha$ values are around 1-1.5 $\times 10^6$ °K and
there is a slight decrease starting at 2.6 R$_\odot$, for both regions 1 and 2; on the other hand the oxygen ion kinetic temperatures show a rapid increase from 2.6 R$_\odot$ in region 1  and from 1.9 R$_\odot$ in region 2. The broadening of the spectral lines can be a signature of energy deposition in the extended corona, which causes the solar wind acceleration, as suggested by the interpretation of coronal hole observations (e.g. Antonucci et al. 2000). Therefore, above 2.6  R$_\odot$, where probably the transition from closed to open magnetic field lines occurs, energy deposition may take place also in the internal parts of the streamer and this is also a possible signature of the current sheet formation.
 We have also derived the electron density for the two regions inside and outside the streamer and the results are shown in the right panel of Figure 3 as a function of heliodistance. The electron density values relative to the inner part of the streamers (region 1, blue dots)
   are compatible with a configuration of static plasma up to 2.6 R$_\odot$ and the values are comparable to those derived in streamers
    by Gibson et~al. (1999)
     (dashed line). From 2.6 to 3.4 R$_\odot$, we obtain values (blue triangle) by assuming a radial expanding coronal plasma achieving outflow velocities between 160 and 180 km/s. This result confirms that at these heights the current sheet is already formed. 
 For what concerns the region external to the streamer (region 2),
we have computed the electron density
taking into account the magnetic topology of the flux tubes as derived by the MHD model
 and the
 results are shown as orange dots in the right panel of Fig. 3. They are intermediate between
those derived for streamers by Gibson et~al.
(1999) (dashed line) and  for coronal holes by Guhathakurta et~al.
(1999) (dotted line). The outflow velocity values found in this region are in the range of 130-240 km/s.
The results of kinetic temperature and electron density values lead to the identification of the sources of the
  slow coronal wind that is found to flow adjacent to
the streamer boundary in the open magnetic field line region. Moreover, the presence of outflowing plasma
  has been accurately detected where the heliospheric current sheet is forming, about at 2.6 R$_\odot$. This
   implies that a contribution to the slow wind comes also from the cusp of
   the streamer above the closed magnetic field lines.
The resulting scenario is  compatible  with the model proposed by
 Wang et~al. (2000), of a  two--component slow wind: one component
 flowing  along the rapidly diverging open magnetic field lines adjacent to
 the streamer boundary, and the second one confined to the region of the
 denser equatorial plasma sheet.

\section{Acknowledgments}
UVCS is a joint project of the
National Aeronautics and Space Administration (NASA), the Agenzia
Spaziale Italiana (ASI) and Swiss Founding Agencies.
The research of LA has been funded through the contract
 I/023/09/0 between the National Institute for Astrophysics (INAF)
 and the Italian Space Agency (ASI) and by Italian Embassy within Egypt Italy Science Year (EISY) 2009 program.

\end{document}